\def\simgt{$_>\atop{^\sim}$}
\def\lya{Ly$\alpha$\ }
\documentstyle[12pt]{article}
\begin{document}

{\center
{\Large {\bf High Redshift Lyman Limit and \\
Damped Lyman-Alpha Absorbers}} \\
\bigskip
{\small L.J. Storrie-Lombardi$^{1,2}$, R.G. McMahon$^{1,5}$, M.J. Irwin$^3$,
C. Hazard$^{1,4}$} \\

\bigskip
{\tiny
{\noindent $^1$Institute of Astronomy, Madingley Road,
Cambridge CB3 0HA  UK \\
$^2$current address: UCSD-CASS, Mail Code 0111,
La Jolla, CA  92093  USA  lsl@ucsd.edu \\
$^3$Royal Greenwich Observatory, Madingley Road, Cambridge CB3 0HE  UK
mike@ast.cam.ac.uk \\
$^4$University of Pittsburgh, Pittsburgh, PA USA} \\
$^5$rgm@ast.cam.ac.uk \\
}
\bigskip
{\small To appear in ``ESO Workshop on QSO Absorption Lines''\\
 Preprint:  astro-ph/yymmddd \\}
}
\begin{abstract}
We have obtained high signal:to:noise optical spectroscopy at 5\AA\
resolution of 27 quasars from the APM z$>$4 quasar survey.  The
spectra have been analyzed to create new samples of high redshift
Lyman-limit and damped Lyman-$\alpha$ absorbers.  These data have been
combined with published data sets in a study of the redshift evolution
and the column density distribution function for absorbers with
$\log$N(HI)$\ge17.5$, over the redshift range 0.01 $<$ z $<$ 5.
The main results are:
\begin{itemize}
\item  Lyman limit systems:
The data are well fit by a power law $N(z) = N_0(1 + z)^{\gamma}$
for the number density per unit redshift.
For the first time intrinsic evolution
is detected
in the product of the absorption cross-section and comoving spatial
number density for an $\Omega = 1$ Universe.
We find $\gamma = 1.55$ ($\gamma = 0.5$ for no evolution)
and $N_0 = 0.27$ with
$>$99.7\% confidence limits for $\gamma$ of 0.82 \& 2.37.
\item Damped \lya systems:
The APM QSOs provide a substantial increase in the redshift path
available for damped surveys for $z>3$.
Eleven candidate and three confirmed damped Ly$\alpha$ absorption systems,
have been identified in the APM QSO spectra covering
the redshift range $2.8\le z \le 4.4$ (11 with $z>3.5$).
Combining the APM survey confirmed and candidate damped \lya
absorbers with previous surveys, we find evidence for a turnover
at z$\sim$3 or a flattening at z$\sim$2 in the cosmological mass density of
neutral gas, $\Omega_g$.
\end{itemize}
The Lyman limit survey results are published in Storrie-Lombardi, et~al.,
1994, ApJ, 427, L13.  Here we
describe the results for the DLA population of absorbers.

\end{abstract}

\section{Introduction}
How and when galaxies formed are questions at the forefront
of work in observational cosmology.
Absorption systems detected in quasar spectra
provide the means to study these phenomena up to z$\sim$5,
back to when the Universe was less than 10\%
of its present age. While the baryonic content of spiral galaxies
that are observed in the present epoch is concentrated in stars,
in the past this must have been in the form of gas.
Damped \lya absorption (DLA) systems have neutral hydrogen column densities of
N(HI)$> 2 \times 10^{20}$cm$^{-2}$.  They dominate the baryonic mass
contributed by HI.  The principal gaseous
component in spirals is HI which has led
to surveys for absorption systems detected by the DLA
they produce
(Wolfe, Turnshek, Smith \& Cohen 1986 [WTSC];
Lanzetta et~al. 1991 [LWTLMH];
Lanzetta, Wolfe \& Turnshek 1995 [LWT95]).
We extend the earlier work on Lyman limit systems and
DLAs to higher redshifts using observations of QSOs from the
APM z$>$4 QSO survey (Irwin, McMahon \& Hazard 1991),
These data more than triple the redshift path surveyed at z$>$3
and allow the first systematic study up to z=4.5.

\section{APM Damped Lyman-Alpha Survey at z$\sim$4}
We have obtained 27 high S/N spectra at 5\AA\
resolution at the William Herschel Telescope.
The spectra were analyzed starting 3000 km s$^{-1}$
blueward of z(emission).
The analysis
was stopped when the S/N ratio became too low
to detect a \lya line with
W(rest)$\ge$5\AA.  This point was
typically caused by the incidence of a Lyman limit system.
Features with W(observed) \simgt 25\AA\ were
selected with an automated procedure.  Most of these are
blends of the dense \lya forest features present at high redshift.
The equivalent width and FWHM were measured interactively
and N(HI) was estimated
for features with W or FWHM $>$30\AA.
Of the 34 measured, 15 have estimated
N(HI)$\ge2\times10^{20}$ cm$^{-2}$ covering $2.8\le z \le 4.4$.
Only one candidate has estimated N(HI)$\ge10^{21}$ cm$^{-2}$.
High resolution spectroscopy of 4 candidates has confirmed 3
as damped (log N(HI)$\ge$20.3).
The sensitivity
of the survey with redshift was determined using the
method developed by Lanzetta (see LWT95). The function
$g(z)$ is calculated,
giving the number of lines of sight along
which a damped system at a redshift $z$ could be detected.
Figure 1a shows the sensitivity of the APM survey alone
and in combination with the WTSC and LWTLMH surveys.

\section{Evolution of the Number Density per Unit Redshift}
The candidate and confirmed DLA systems from
the APM sample and previous surveys
(WTSC; LWTLMH; LWT95)
have been combined to study the evolution of the
number density per unit redshift for 0.01 $<$ z $<$ 4.7.
Fit with the customary power law $N(z)=N_0(1+z)^{\gamma}$,
a population with no intrinsic evolution
in the product of the absorption cross-section and comoving spatial
number density will have $\gamma=1/2$ ($\Omega = 1$)
or $\gamma=1$ ($\Omega=0$).
A maximum likelihood fit to the data with z$>$1.5 yields
$N(z)=0.03(1+z)^{1.5\pm0.6}$, consistent with
no intrinsic evolution
even though the value of $\gamma$ is similar
to that found for the Lyman limit
systems where evolution is detected at a significant level.
However, there is redshift evolution evident in the
higher column density systems with an apparent decline in $N(z)$
for z$>$3.5.
These results are displayed in figure 1(b).
The combined data set is plotted as dashed lines with the
above fit.  The results for only the absorbers with
log N(HI)$\ge21$ are shown as solid lines.
The z$<$1.5 bin is taken from LWT95.

\section{Evolution of $\Omega_g$ -- Baryons in Neutral Gas}
The mean cosmological mass density contributed by damped \lya
absorbers can be estimated as
$$ \langle \Omega_g \rangle = {H_0 \mu m_H \over c \rho_{crit}}
\int_{N_{min}}^\infty Nf(N)dN $$
as defined in LWTLMH (equations 17-18),
giving the current mass density in units of the current
critical density. The errors in $\Omega_g$ are
difficult to estimate because the column density distribution
function, $f(N)$, is not known.  LWTLMH utilised the standard error
in the distribution of N(HI)
which yields zero error if all the column densities in a bin are
the same.
We have estimated the fractional variance in $\Omega_g$ as
$ \sum_{i=1}^p N_i^2 / (\sum_{i=1}^p N_i)^2  $
which yields $\sqrt{n}$ errors if all the column densities included
in a bin are equal. This method yields larger errors.
The results for $\Omega_g$ are shown in figure 2 for q$_0$=0
and q$_0$=0.5 (H$_0$=50).  The z$<$1.5 bin
is taken from LWT95.  The z$>$1.5 solid bins utilise
the data from WTSC, LWTLMH, and the APM survey.  The dotted bins
exclude the APM data.
The inclusion of the APM survey data for $z>3$ lowers the
value previously found for $3 < z < 3.5$ and indicates a possible
turnover for $z > 3.5$.  The results are also consistent with
a relatively constant value of $\Omega_g$ for $z>2$ as the
error bars are still very large at high redshift.
Larger samples of bright $z>4$ quasars are needed.

\section{Summary}
The QSOs from the APM survey more than triple the $z>3$ redshift path
for DLA surveys.
Fourteen candidate DLA systems
have been identified in the APM spectra covering
$2.8\le z \le 4.4$ (11 with $z>3.5$), with 3 confirmed.
Combining these data with the previous surveys
and fitting a single power law for z$>$1.5
gives N(z)$=.03(1+z)^{1.5\pm0.6}$, marginally
consistent with no evolution models.
Evolution is evident in the highest
column density absorbers with the incidence of
systems with log N(HI)$\ge$21
apparently decreasing for z\simgt 3.5. We find
evidence for a turnover or flattening in the
cosmological mass density of neutral gas, $\Omega_g$
at high redshift.  The more gradual evolution of $\Omega_g$ than
previously found helps alleviate the `cosmic G-dwarf problem' (LWT95), i.e.
if a large amount of star formation has taken place between
z=3.5 and z=2, a much larger percentage of low metallicity stars
should exist than is detected.  It is also consistent with the suggestion
by Pettini et~al. (1994) that the wide range in DLA metallicities measured
at the same epoch indicates that at z$\sim$2 they are observed prior to
the bulk of star formation in the disk.

\vfill\eject

\bigskip
\bigskip
{\Large\noindent{\bf Figure Captions}}
\bigskip

\noindent {\bf Figure 1}:  (a) The sensitivity function, $g(z)$, of the
DLA surveys.  This gives
the number of lines of sight along which a
damped system at redshift $z$ could be detected.
(b) The number density of DLA per unit redshift,
$N(z)$, vs. z(absorption). The dashed bins show
N(z) for all the damped systems and the solid bins
for systems with N(HI)$\ge10^{21}$ cm$^{-2}$.
A single power law fit to the sample for z$>$1.5
gives N(z)$=.03(1+z)^{1.5\pm0.6}$.

\bigskip
\noindent {\bf Figure 2}: The mean cosmological mass density
in neutral gas, $\Omega_g$,
contributed by DLA absorbers for 0.01$\le$z$\le$4.7
for q$_0$=0 and q$_0$=0.5 (H$_0$=50).  The z$<$1.5 bin
is taken from LWT95.  The z$>$1.5 solid bins utilise
the WTSC, LWTLMH, and APM survey data. The dotted bins
exclude the APM data. The points at z$=$0 are $\Omega_{stars}$
(star) from Gnedin \& Ostriker (1992) and
$\Omega_{HI}$ (circle) from Rao \& Briggs (1993).

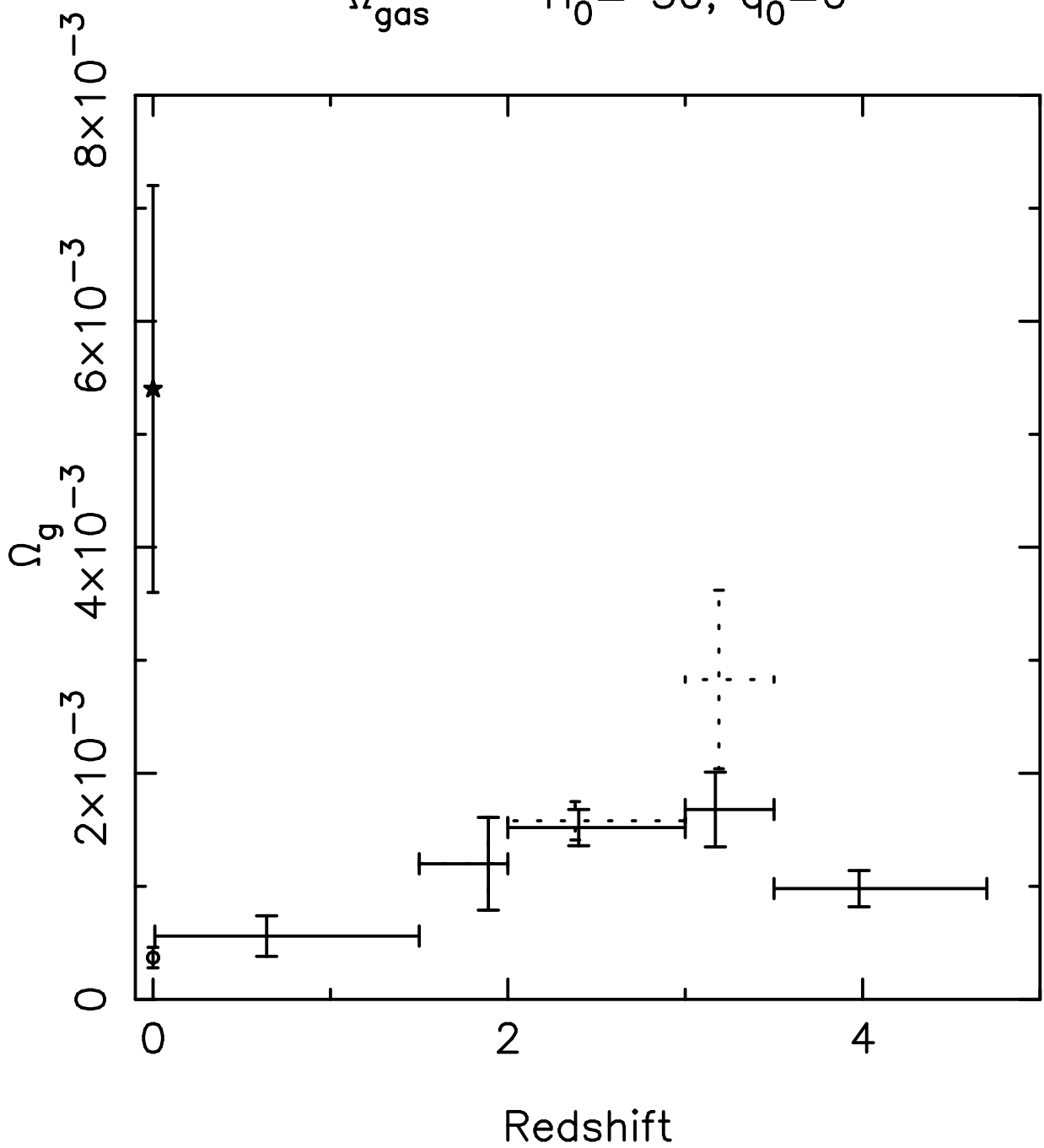
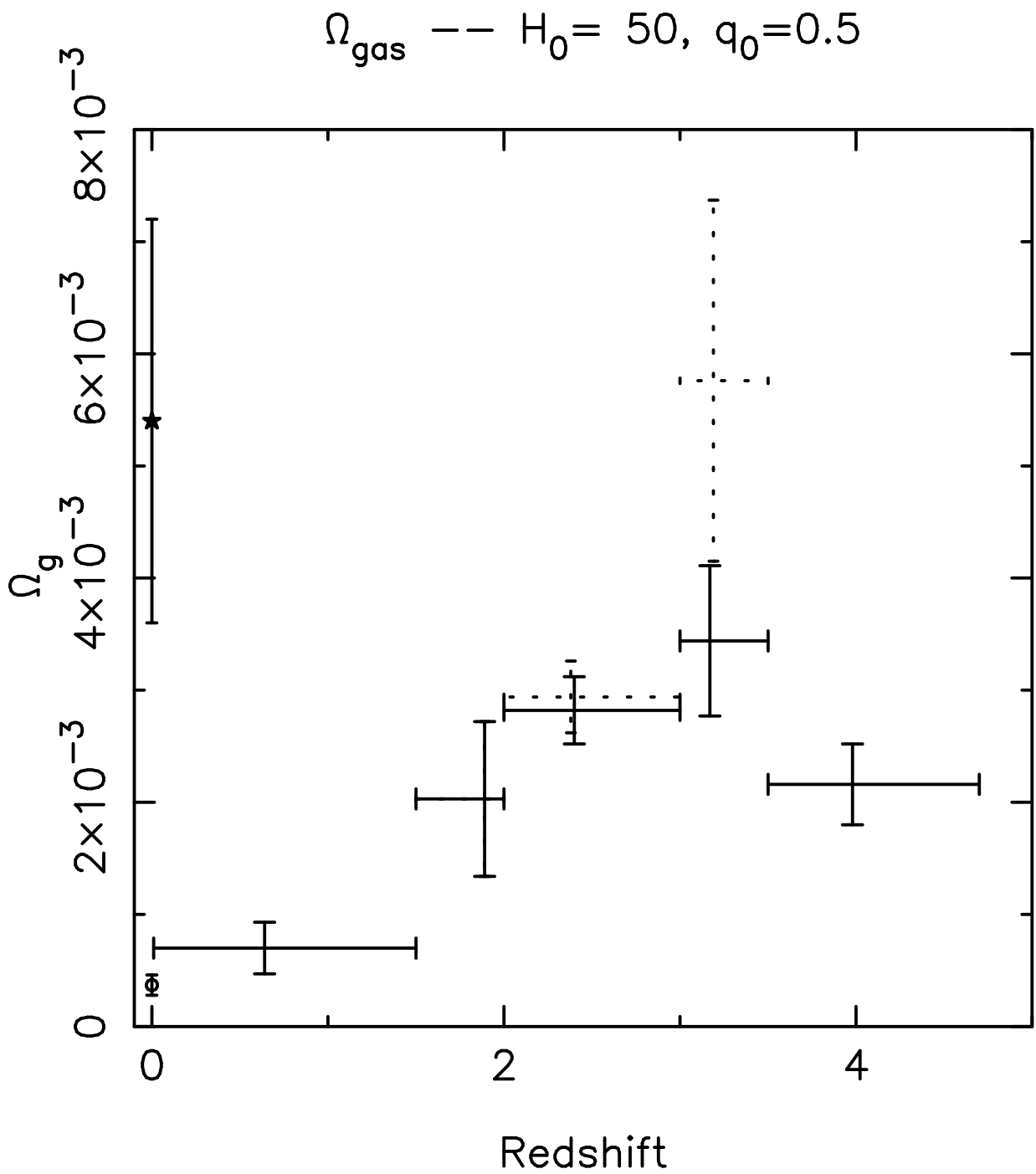
\end{document}